\documentclass[12pt]{article}
\usepackage {mathtext}
\usepackage[dvips]{graphicx}
\newcommand{\be}{\begin{eqnarray} }
\newcommand{\ee}{\end{eqnarray} }
\newcommand{\beq}{\begin{equation} }
\newcommand{\eeq}{\end{equation} }

\begin{document}

\begin{center}{\large\bf
Polarized parton distributions from NLO QCD analysis of world DIS and SIDIS data
 \vskip 1cm
\rm
{A. Sissakian\footnote{email: sisakian@jinr.ru}, O. Shevchenko\footnote{email: shev@mail.cern.ch}, O. Ivanov\footnote{email: ivon@jinr.ru}\\
{\it  Joint Institute for Nuclear Research, 141980 Dubna, Russia}\\
}
}
\end{center}
\begin{abstract}
The combined analysis of polarized DIS and SIDIS data is performed in NLO QCD.
The new parametrization on polarized PDFs is constructed. The uncertainties on
PDFs and their first moments are estimated applying the modified Hessian method.
 The especial attention
is paid to the impact of novel SIDIS data on the polarized distributions of light sea and strange
quarks. In particular, the important question of polarized sea symmetry is studied in comparison
with the latest results on this subject.

\end{abstract}
\begin{flushleft}
\end{flushleft}

Since the observation of the famous spin crisis in 1987 \cite{EMC_crisis} one of the most
intriguing and still unsolved problems of the modern high energy physics is
the nucleon spin puzzle. The key component of this problem, which attracted
the great both theoretical and experimental efforts during many years is
the finding of the polarized parton distributions functions (PDFs) in nucleon.

The analysis of data on inclusive polarized DIS
enables us to extract such important quantities as the singlet $\Delta\Sigma(x,Q^2)$
and nonsinglet $\Delta q_3(x,Q^2)$, $\Delta q_8(x,Q^2)$ combinations of the polarized PDFs, 
and, thereby, the sums of valence and sea PDFs
$\Delta q+\Delta\bar q\equiv\Delta q_V+2\Delta\bar q$.
Besides, dealing with DIS data, the gluon helicity distribution $\Delta G(x,Q^2)$ is
determined due to the evolution in singlet sector and weak dependence on $\Delta G$ of the
polarized structure function $g_1$ in NLO QCD. 
However, even for singlet combinations $\Delta u+\Delta\bar u$, $\Delta d+\Delta\bar d$,
$\Delta s+\Delta\bar s$, considered as well-determined within DIS, we still meet the problem:
there are only two equations corresponding to inclusive asymmetries $A_1$
measured on proton and deuteron targets from which we should determine
three unknown combinations $\Delta q_3(x)$, $\Delta q_8(x)$ and  $\Delta\Sigma(x)$
(or, alternatively, $\Delta u+\Delta\bar u$, $\Delta d+\Delta\bar d$,
$\Delta s+\Delta\bar s$).
So, it is unavoidable to involve some additional assumptions performing the 
fitting procedure for the purely inclusive DIS data. Moreover, DIS data can not help
us to solve the important problem of valence and sea PDFs separation.

The basic\footnote{As long as no neutrino factory is built and no hyperdense polarized
target is created, we cannot study the DIS processes with the
neutrino beam, which would enable us to find the polarized
valence and sea quark distributions separately.} process which enables us to solve these 
problems is the process of semi-inclusive DIS (SIDIS). 
However, until recently the quality of the polarized SIDIS data was rather poor,
so that its inclusion in the analysis did not helped us \cite{florian00} to solve the main task
of SIDIS measurements: to extract the polarized 
sea and valence PDFs of all active flavors. Only in 2004 the first polarized SIDIS
data with the identification of produced hadrons (pions and kaons) were published \cite{hermes04}. 
These data were included in the global QCD analysis in Ref. \cite{florian05}. Further, COMPASS \cite{compass_hpm}
presented the polarized SIDIS data (without identification of hadrons in the final state), in particular, in the 
low $x$ region unaccessible for HERMES. This data were included in the latest parametrization 
of Ref. \cite{dssv08}.   Recently, the new data on the 
SIDIS asymmetries $A_d^{\pi^\pm}, A_d^{K^\pm}$ were published \cite{compass09} by the COMPASS
collaboration. It is of importance that this data cover the most important and badly 
investigated low $x$ region. In this paper we include  this data in the new global QCD analysis
of all existing polarized DIS and SIDIS data. The elaborated parametrization on the polarized
PDFs in some essential points 
differs from the parametrization of Ref. \cite{dssv08} (see below).
Nevertheless, the results on PDFs obtained with both parametrizations are compatible within
the errors.

Constructing the new parametrization we tried to be as close as possible to our previous NLO QCD
analysis of pure inclusive DIS data \cite{compass_dis}. Namely, to properly describe the DIS
data we (just as before \cite{compass_dis}) parametrize the singlet $\Delta\Sigma$ and two
nonsinglet $\Delta q_3$, $\Delta q_8$ combinations at the initial scale $Q_0^2=1\,GeV^2$
\be
\label{param_sigma}
\Delta\Sigma=\eta_{\Delta\Sigma}  \frac{ x^{\alpha_{\Delta\Sigma}} \,(1-x)^{\beta_{\Delta\Sigma}} }{\int_{0}^{1} x^{\alpha_{\Delta\Sigma}} \,(1-x)^{\beta_{\Delta\Sigma}}  dx},\\
\label{param_q3}
\Delta q_3=\eta_{\Delta q_3}  \frac{ x^{\alpha_{\Delta q_3}} \,(1-x)^{\beta_{\Delta q_3}} \,(1+\gamma_{\Delta q_3} x)}{\int_{0}^{1} x^{\alpha_{\Delta q_3}} \,(1-x)^{\beta_{\Delta q_3}} \,(1+\gamma_{\Delta q_3} x) dx},\\
\label{param_q8}
\Delta q_8=\eta_{\Delta q_8}  \frac{ x^{\alpha_{\Delta q_8}} \,(1-x)^{\beta_{\Delta q_8}} \,(1+\gamma_{\Delta q_8} x+\delta_{\Delta q_8} \sqrt{x})}{\int_{0}^{1} x^{\alpha_{\Delta q_8}} \,(1-x)^{\beta_{\Delta q_8}} \,(1+\gamma_{\Delta q_8} x+\delta_{\Delta q_8}\sqrt{x}) dx}.
\ee
Then, the quantities $\Delta u+\Delta\bar u$, $\Delta d+\Delta\bar d$, $\Delta s=\Delta\bar s$ are determined
as:
\be
\label{param_uub}
&&\Delta u+\Delta\bar u=\frac{1}{6}(3\Delta q_3+\Delta q_8+2 \Delta\Sigma),\quad
\Delta d+\Delta\bar d=\frac{1}{6}(3\Delta q_3-\Delta q_8+2 \Delta\Sigma),\\
&&\Delta s=\Delta\bar s=\frac{1}{6}(\Delta\Sigma-\Delta q_8).
\ee
The gluon PDF is parametrized as 
\be
\label{param_gl}
\Delta G=\eta_{\Delta G} \frac{ x^{\alpha_{\Delta G}} \,(1-x)^{\beta_{\Delta G}} \,(1+\gamma_{\Delta G} x)}{\int_{0}^{1} x^{\alpha_{\Delta G}} \,(1-x)^{\beta_{\Delta G}} \,(1+\gamma_{\Delta G} x) dx}.
\ee
It is easy to see that in our parametrization the coefficients $\eta$ are just the first moments
of the respective local quantities. In particular, 
the advantage of $\Delta q_3$ and $\Delta q_8$
parametrization  in the form (\ref{param_q3}), (\ref{param_q8}) is that with the such choice it is very convenient
to apply and control the $SU_f(2)$ and $SU_f(3)$ sum rules 
\be
\label{eq3}
\eta_{\Delta q_3}\equiv \Delta_1 q_3\equiv \int_0^1 dx \Delta q_3(x)=F+D=1.269\pm0.003,\\
\label{eq8}
\eta_{\Delta q_8}\equiv \Delta_1 q_8\equiv \int_0^1 dx \Delta q_8(x)=3F-D=0.586\pm0.031.
\ee
Besides, the such inheritance with Ref. \cite{compass_dis} allow us to clearly see 
the impact of SIDIS data on the results of pure inclusive DIS data analysis.

Further, to properly describe the SIDIS data we, besides $\Delta\Sigma$, 
$\Delta q_3$ and $\Delta q_8$,  parametrize the sea PDFs of $u$ and $d$ flavors:
\be
\label{param_ub}
\Delta\bar u=\eta_{\Delta\bar u}  \frac{ x^{\alpha_{\Delta\bar u}} \,(1-x)^{\beta_{\Delta\bar u}} }{\int_{0}^{1} x^{\alpha_{\Delta\bar u}} \,(1-x)^{\beta_{\Delta\bar u}}  dx},\quad
\Delta\bar d=\eta_{\Delta\bar d}  \frac{ x^{\alpha_{\Delta\bar d}} \,(1-x)^{\beta_{\Delta\bar d}} }{\int_{0}^{1} x^{\alpha_{\Delta\bar d}} \,(1-x)^{\beta_{\Delta\bar d}}  dx}.
\ee
Then, $\Delta u$ and $\Delta d$ are determined from Eqs. (\ref{param_uub}) and (\ref{param_ub}),
while the valence PDFs are determined by $\Delta u_V=\Delta u-\Delta\bar u$ and  $\Delta d_V=\Delta d-\Delta\bar d$.
Thus, all polarized PDFs are completely determined within the parametrization.

Comparing DIS sector, Eqs. (\ref{param_sigma})-({\ref{param_q8}}), (\ref{param_gl}) with the
respective  parametrizations from Ref. \cite{compass_dis},
one can see some distinctions. These are additional factors $\gamma_{\Delta q_3} x$, $\gamma_{\Delta q_8} x$ 
in Eqs. (\ref{param_q3}), (\ref{param_q8}) which are introduced
to provide the better flexibility of the parametrizations on the respective quantities, required by the inclusion
of SIDIS data. 
Besides, we introduce the additional factors $\delta_{\Delta q_8} \sqrt{x}$
in Eq. (\ref{param_q8})  and $\gamma_{\Delta G} x$ in Eq. (\ref{param_gl})
to provide the possibility of sign-changing scenarios for $\Delta s$ and $\Delta G$, respectively (see below).
We consider two options for handling of $SU_f(2)$ and $SU_f(3)$ sum rules. In the first case
we apply the constraints (\ref{eq3}) and (\ref{eq8}) putting $\eta_{\Delta q_3}$ and $\eta_{\Delta q_8}$
equal to the central values of the respective constants. In the second case we allow 
 $\eta_{\Delta q_3}$ and $\eta_{\Delta q_8}$ to vary within the uncertainties for $F+D$ and $3F-D$. 
It is of importance that, as we will see below, both options produce almost the same results.

We analyze the inclusive $A_1$ and semi-inclusive $A_1^h$ asymmetries.
The inclusive asymmetry reads
\be
\label{a_1}
A_{1p}=\frac{g_{1p}(x,Q^2)}{F_{2p}(x,Q^2)/\{2x(1+R(x,Q^2))\}},
\ee
where the polarized structure function $g_{1p}$ in NLO QCD looks as
\be
2g_{1p}=\sum_{q,\bar q}\left\{\Delta q+\frac{\alpha_s(Q^2)}{2\pi}[\Delta C_q\otimes \Delta q+\frac{1}{f}\Delta C_g\otimes\Delta G]\right\}.
\ee
Throughout this paper we use the $\overline{MS}$ factorization scheme.
The respective coefficient functions $\Delta C_{q,g}$ can be found in Ref. \cite{Cq}. The semi-inclusive asymmetries,
besides $x$ and $Q^2$, depend also on hadronic variable $z$. As usual, we apply the semi-inclusive asymmetries
\be
\label{a_1h}
A_{1p}^h(x,Q^2)=\frac{\int_{0.2}^{1}dz g_{1p}^h(x,Q^2,z)}{\int_{0.2}^{1}dzF_{2p}^h(x,Q^2,z)/\{2x(1+R(x,Q^2))\}}
\ee
integrated over the cut $z>0.2$, which corresponds to the current fragmentation region.
The semi-inclusive structure functions $g_1^h$ in NLO QCD are given by 
\be
2g_{1p}^h&=&\sum_{q,\bar{q}} e_q^2\Delta q[1+\otimes
 \frac{\alpha_s}{2\pi}\Delta C_{qq}\otimes]D^h_{q}
\nonumber+(\sum_{q,\bar{q}} e_q^2\Delta q)\otimes \frac{\alpha_s}
{2\pi}\Delta C_{gq}\otimes D^h_g\nonumber\\&+&\Delta G\otimes
 \frac{\alpha_s}{2\pi}\Delta C_{qg}\otimes(\sum_{q,\bar{q}} e_q^2
D^h_{q}).
\ee
The respective Wilson coefficients $\Delta C_{qq,qg,gq}$ can be found in \cite{florian-formulas}.  Of great importance
for the SIDIS data analysis is the choice of parametrization on the fragmentation functions.
We use here the latest NLO parametrization from Ref. \cite{dss}. Calculating $F_2$ in Eq. (\ref{a_1})
and $F_2^h$ in Eq. (\ref{a_1h}) we use parametrization for $R$ from \cite{r_param}
and the  recent NLO parametrization on unpolarized PDFs from Ref. \cite{GJR}.
For the $\alpha_s(Q^2)$ calculation
we apply the same procedure as in Ref. \cite{GJR} (i.e. $\alpha_s(Q^2)$ in $\overline{MS}$ scheme
is calculated just as in \cite{GJR} with $\alpha_s(M_Z^2)=0.1145$).
The deuteron structure functions $g_{1d}$ and $g_{1d}^h$ are calculated applying 
$g_{1d}^{(h)}=(g_{1p}^{(h)}+g_{1n}^{(h)})(1-\frac{3}{2}\omega_d)/2$ with $\omega_d=0.058$ (see, for instance \cite{compass_dis}). 
In our analysis the positivity constraint $|\Delta q|<q$ and $|\Delta G|<G$ holds with the precision $0.001$.

Of importance is the correct application of the 
factor $1+\gamma^2=1+4M^2x^2/Q^2$ in the analysis -- see the discussion on this
question in Ref. \cite{sidorov}. When one rewrites the asymmetry $A_1$ in terms of $F_2$
instead of $F_1$ then this factor precisely cancel out 
in the ratio of $g_{1}(1+\gamma^2)$ and $F_1=F_2(1+\gamma^2)/2x(1+R)$, 
so that we just arrive at right-hand side of Eq. (\ref{a_1}).
The same cancellation holds in SIDIS case, so that one arrives at Eq. ({\ref{a_1h}}).
At the same time, sometimes \cite{clas} the inclusive data is tabulated in the form
$g_1/F_1$ (not in the form $A_1=(1+\gamma^2)g_1/F_1$). In this case 
we fit the experimental values $g_1/F_1$ by the $g_1/[(1+\gamma^2)F_2/2x(1+R)]$.

One of the main conditions of the successful global QCD
analysis is the robust program for the DGLAP solution. 
The respective program should be fast, well tested and should provide a good precision of DGLAP solution.
The elaborated in Ref. \cite{compass_dis} program for the polarized DIS data analysis,
based on inverse Mellin transformation method,
satisfies to all these requirements. So, 
we again use here this program package, properly modifying it
in accordance with the peculiarities of SIDIS data (calculation of SIDIS structure functions
and asymmetries in the space of Mellin moments are included).

All procedures of the global QCD analysis are based on the construction of the effective $\chi^2$ function
that describes the quality of the fit to data for a given set of varying parameters $\{a_i\}$.
In the case of polarized DIS and SIDIS analysis one usually uses the $\chi^2$ function in the form
(see, for instance, \cite{dssv08} for detail)
\be
\label{chi2}
\chi^2=\sum_{i}\left(\frac{A_{exp}-A_{theor}(\{a_i\})}{\delta (A_{exp})}\right)^2,
\ee
where  $A_{exp}$ is the measured value of the asymmetry, $\delta(A_{exp})$ is its uncertainty\footnote{We treat 
$(\delta(A_{exp}))^2$ in Eq. (\ref{chi2})  as the quadratic sum of the statistical
and systematical errors (see discussion on this question in Ref. \cite{dssv08}).},
$A_{theor}$ is its theoretical estimation. 
For the minimization of $\chi^2$ function we use the MINUIT package \cite{minuit}.

For our analysis we collected all accessible in literature polarized DIS and SIDIS data.
We include the inclusive proton data from Refs \cite{smc,e143,E155_d,HERMES,clas},
inclusive deuteron data from Refs  \cite{emc,smc,e143,e155p,HERMES,clas} 
and inclusive neutron data from Refs  \cite{e142,e154,jlab,hrm1}. The semi-inclusive data
(asymmetries $A_{p,d}^{h^+,h^-}$, $A_{p,d}^{\pi^+,\pi^-}$, $A_{d}^{K^+,K^-}$) are taken
from Refs \cite{hermes04,smc_semi,compass_hpm} and, besides, we include the latest
COMPASS data from Ref \cite{compass09}. In total we have 232 points for the inclusive polarized DIS
and 202 points for semi-inclusive polarized DIS. 
If we fix $\eta_{\Delta q_3}$ and $\eta_{\Delta q_8}$ by the center values of $F+D$ and $3F-D$
in sum rules (\ref{eq3}), (\ref{eq8}), then for 16 fit parameters
 $\chi_0^2|_{inclusive}=188.4$ and
$\chi_0^2|_{semi-inclusive}=194.8$  for DIS and SIDIS data, while $\chi_0^2|_{total}=383.9$ for the full set of
data (434 points).  On the other hand, if $\eta_{\Delta_{q3}}$ and $\eta_{\Delta_{q_8}}$ are varied within
the errors on $F+D$ and $3F-D$, then 
 for 18 fit parameters the resulting $\chi^2$ values are: $\chi_0^2|_{inclusive}=188.2$,
$\chi_0^2|_{semi_inclusive}=194.8$ with $\chi_0^2|_{total}=383.7$.
Thus, one can conclude that the fit quality is quite good: $\chi_0^2/D.O.F.\simeq 0.84$.

The optimal values of our fit parameters are presented in Table \ref{table:parameters}
for both options, with fixed and varied $\eta_{\Delta q_3}$ and $\eta_{\Delta q_8}$.
As it is seen from Table \ref{table:parameters},
the results are almost the same in both cases: the differences in parameters are less than 1\%.
\begin{table}[h!]
\caption{Optimal values of the global fit parameters at the initial scale $Q_0^2=1\,GeV^2$. }
\begin{tabular}{cccc}
  \hline
Parameter&{$\Delta\Sigma$}&$\Delta q_3$&$\Delta q_8$\\
  \hline
  $\alpha$  &    1.0227 (1.0216)& -0.6342 (-0.6380) & -0.7916  (-0.7827)  \\
  $\beta$   &    3.3891 (3.3873)&  3.1418  (3.1398)   &  $=\beta_{\Delta_{q_3}}$ \\   
  $\gamma$  &    0.0 (fixed)    &    23.9180  (24.2240) &   36.8400 (37.1990)  \\
  $\delta$  &    0.0 (fixed)    &    0.0 (fixed)   &   -13.7480 (-13.8390) \\
  $\eta$&    0.3846 (0.3850)        &      1.2660 (1.2690)&   0.6170 (0.5860)\\
\hline
\end{tabular}
\begin{tabular}{ccccc}
\hline
Parameter&{$\Delta G$}&$\Delta\bar u$&$\Delta\bar d$\\
  \hline                
  $\alpha$    &  0.9040  (0.9154)   &   -0.3506 (-0.3412)  &   0.2802  (0.2852)   \\      
  $\beta$    &   $=\beta_{\Delta\bar u}$ &  15.0 (fixed)          &   $=\beta_{\Delta\bar u}$    \\    
  $\gamma$    &  -5.6703 (-5.6317)    &     0.0 (fixed)     &    0.0 (fixed)\\                
  $\delta$    &   0.0000 (fixed)      &     0.0 (fixed)     &    0.0 (fixed)\\                
  $\eta$      &    -0.1828 (-0.1813 )&     0.0672 (0.0670) &    -0.0792    (-0.0794)  \\
\hline
\end{tabular}
\label{table:parameters}
\end{table}

Our calculations show that the fit quality does not decrease if we cancel the extra
parameters setting  $\beta_{\Delta q_8}$ equal to  $\beta_{\Delta q_3}$, since their values occur
very close to each other when we try to find the best fit. Besides, we
use the equality $\beta_{\Delta\bar u}=\beta_{\Delta\bar d}=\beta_{\Delta G}$ 
(just as in Ref. \cite{dssv08}) since
the polarized data at $x>0.6\div0.7$ is just absent, while
in the region $0.4\div0.7$ the statistical errors are too large to feel the difference
in values of these parameters.

Certainly, the construction of the best fit should be accompanied by the reliable method
of uncertainties estimation. We choose the modified Hessian method \cite{pump1}, \cite{mstw} which 
well works (as well as the Lagrange multipliers method -- see \cite{dssv08} and references therein) 
even in the case of deviation of $\chi^2$ profile from the quadratic parabola, 
and was successfully applied in a lot of physical tasks.

\begin{figure}
\includegraphics[height=6cm]{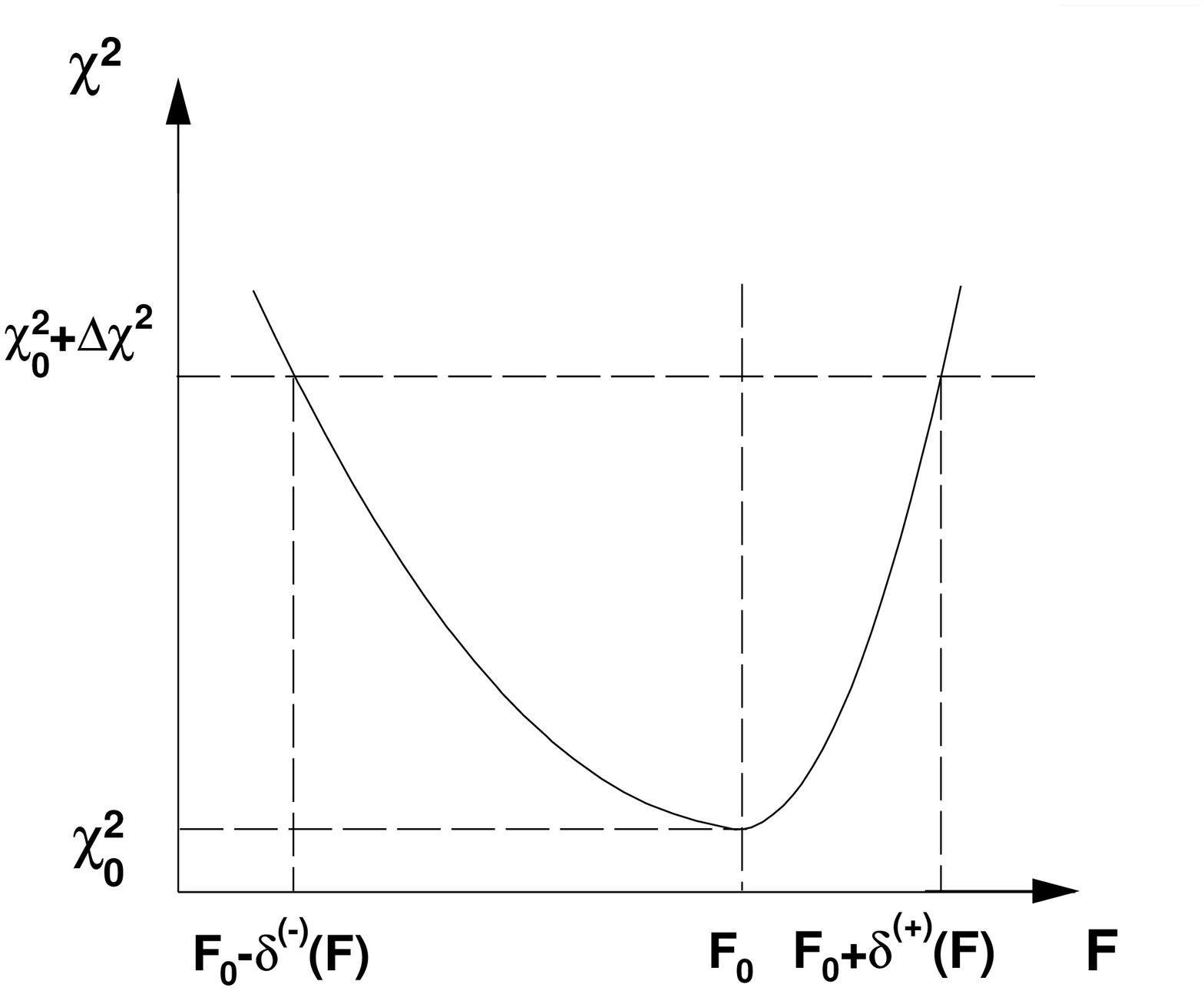}
\caption{The scheme illustrating the search of asymmetric errors within the modified Hessian method.}
\label{chi2_profile}
\end{figure}

Let us recall that the standard Hessian method is based on the assumption that the global $\chi^2$ is quadratic 
near the minimum $\chi_0^2$:

\be
\label{chi2_error}
\chi^2=\chi_0^2+\sum_{i,j=1}^M H_{ij} (a_i-a_i^0)(a_j-a_j^0).
\ee
Here $a_i^0$ are the fit parameters values at the global minimum and $H_{ij}$ are the elements
of the Hessian matrix -- the matrix of second derivatives of $\chi^2$ in the minimum. Then, the uncertainty 
on any physical quantity $F$ which depends on $M$ fit parameters $\{a_i\}$, can be estimated applying 
\be
\label{f_er}
(\delta F)^2=\Delta\chi^2 \sum_{i,j=1}^M\frac{\partial F}{\partial a_i}H^{-1}_{ij}\frac{\partial F}{\partial a_j}.
\ee
 The Hessian method in the simple form (\ref{f_er}) 
is implemented in the MINUIT program \cite{minuit} ("Hesse procedure") and it perfectly works
if $\chi^2$ has the parabolic profile near the minimum. However, in practice we often
meet the problems which decrease the reliability of this simple version of Hessian method,
and main of them is the deviation of  $\chi^2$ from parabolic form -- see \cite{pump1} for
detail. To deal with asymmetric $\chi^2$ profiles 
we apply the modification \cite{pump1}, \cite{mstw} of the standard Hessian method,
which is proved to be well working even in the strongly asymmetric cases.
Shortly, the essence of procedure is following. First,
one starts with still symmetric errors $\pm\delta F$ on the physical
quantity $F$, which one finds via Eq. (\ref{f_er}) rewritten \cite{pump1} in terms
of  eigenvectors and eigenvalues of the diagonalized Hessian matrix\footnote{For
the respective calculations we apply the program ITERATE by J. Pumplin \cite{pump2}.}. Simultaneously,
one calculates the values of fit parameters $\{a_i\}$
corresponding to $F+\delta(F)$ and $F-\delta(F)$ and, thereby, the respective $\chi^2$ values.
To obtain asymmetric errors \cite{mstw} corresponding to the real $\chi^2$ profile one
varies $\Delta\chi^2$ in Eq. (\ref{f_er}) (rewritten in terms of eigenvectors and eigenvalues) 
and calculates the respective
values of the parameters, finding the intersections
of $\chi^2(F)$ curve with the straight line $\chi^2=\chi_0^2+\Delta\chi^2$, where
$\Delta\chi^2$ is already fixed quantity ($\Delta\chi^2=1$ or $\Delta\chi^2=18.065$ here -- see below). 
Then the differences of $F$ values in 
these intersections with $F=F_0$ in the global minimum just determine 
$F+\delta^{(+)}(F)$ and $F-\delta^{(-)}(F)$, where
$\delta^{(\pm)}(F)$ are in general asymmetric uncertainties on the physical quantity $F$ -- see Fig. \ref{chi2_profile}.

Besides, very important question arises about  choice of $\Delta\chi^2$ determining 
the uncertainty size.  The standard choice is
$\Delta\chi^2=1$, 
just as 
we did before in Ref. \cite{compass_dis} (see also \cite{dssv08} and references therein). 
However, the such choice of $\Delta\chi^2$
can lead to underestimation of uncertainties, as it was argued in Ref. \cite{aac08}.
The alternative choice of $\Delta\chi^2$ (see, for example, Refs. \cite{aac08}, \cite{ansel09} 
and references therein) is based  on the equation
\be
\label{probability}
 P = 0.68= \int_0^{\Delta \chi^2}\frac{1}{2\Gamma(M/2)}
\left(\frac{x}{2}\right)^{(M/2)-1}
\exp\left(-\frac{x}{2}\right){\rm d}x,
\ee                    
where P=0.68 ($1\sigma$ deviation) is the probability  to find the values of all $M$ fitting parameters
inside the hypervolume determined
by the condition $\chi^2\le \chi_0^2+\Delta\chi^2$. In our case (17 parameters) the $\Delta\chi^2$ value calculated
from Eq. (\ref{probability}) is equal to $18.065$. We calculate the uncertainties for both 
$\Delta\chi^2=1$ and $\Delta\chi^2=18.065$ options.

\begin{figure}[h!]
\begin{tabular}{cc}
\includegraphics[height=5cm]{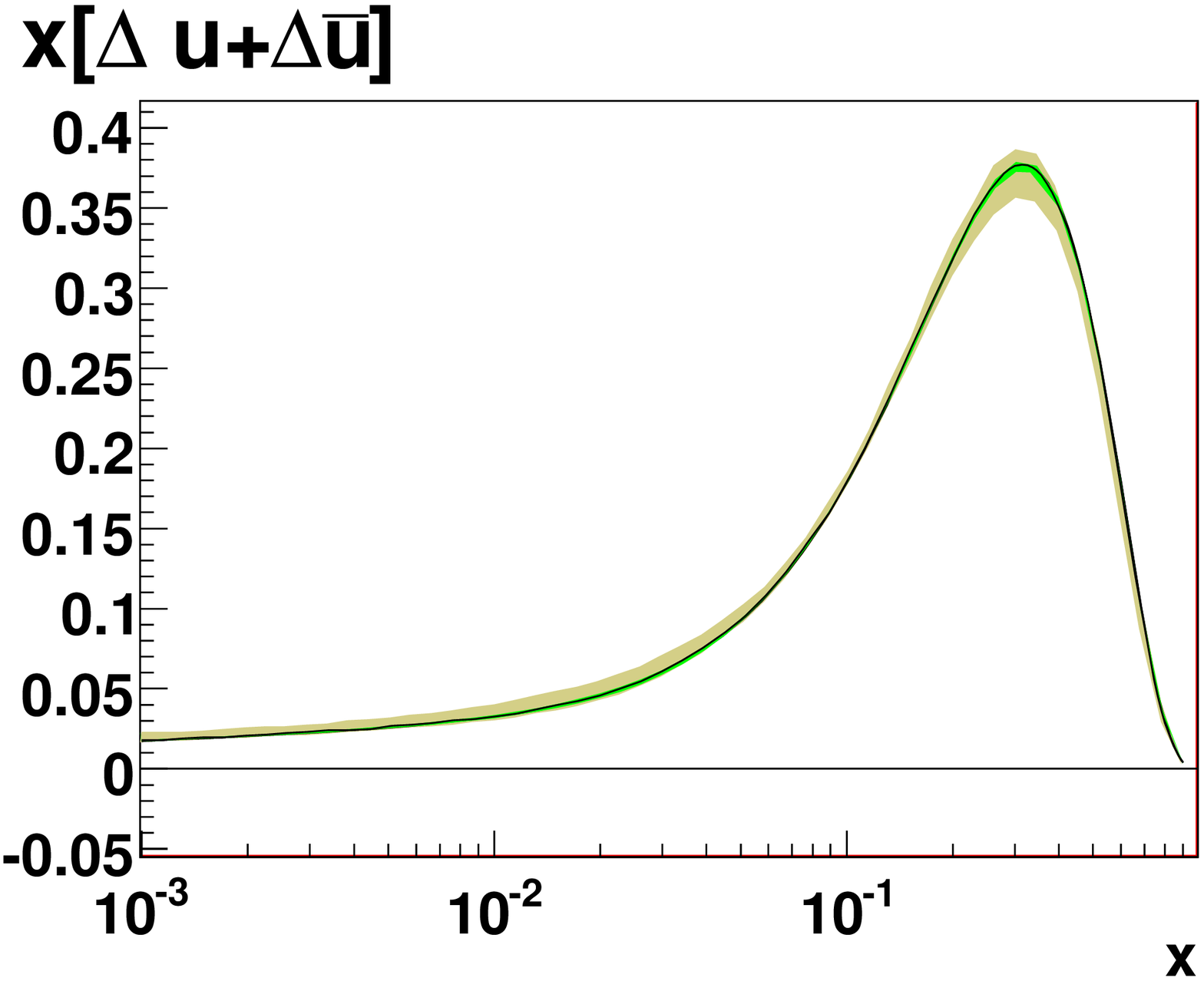}&
\includegraphics[height=5cm]{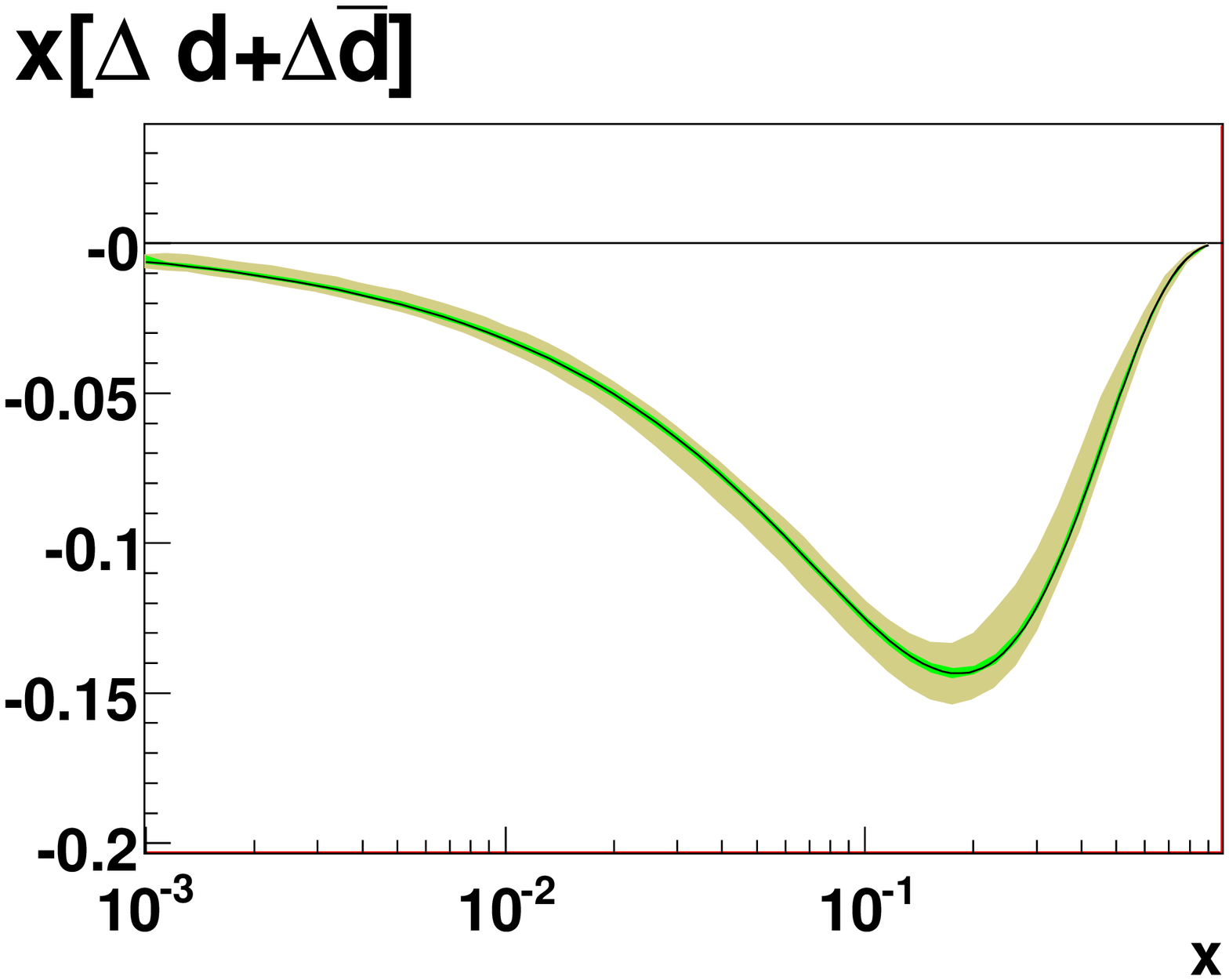}\\
\includegraphics[height=5cm]{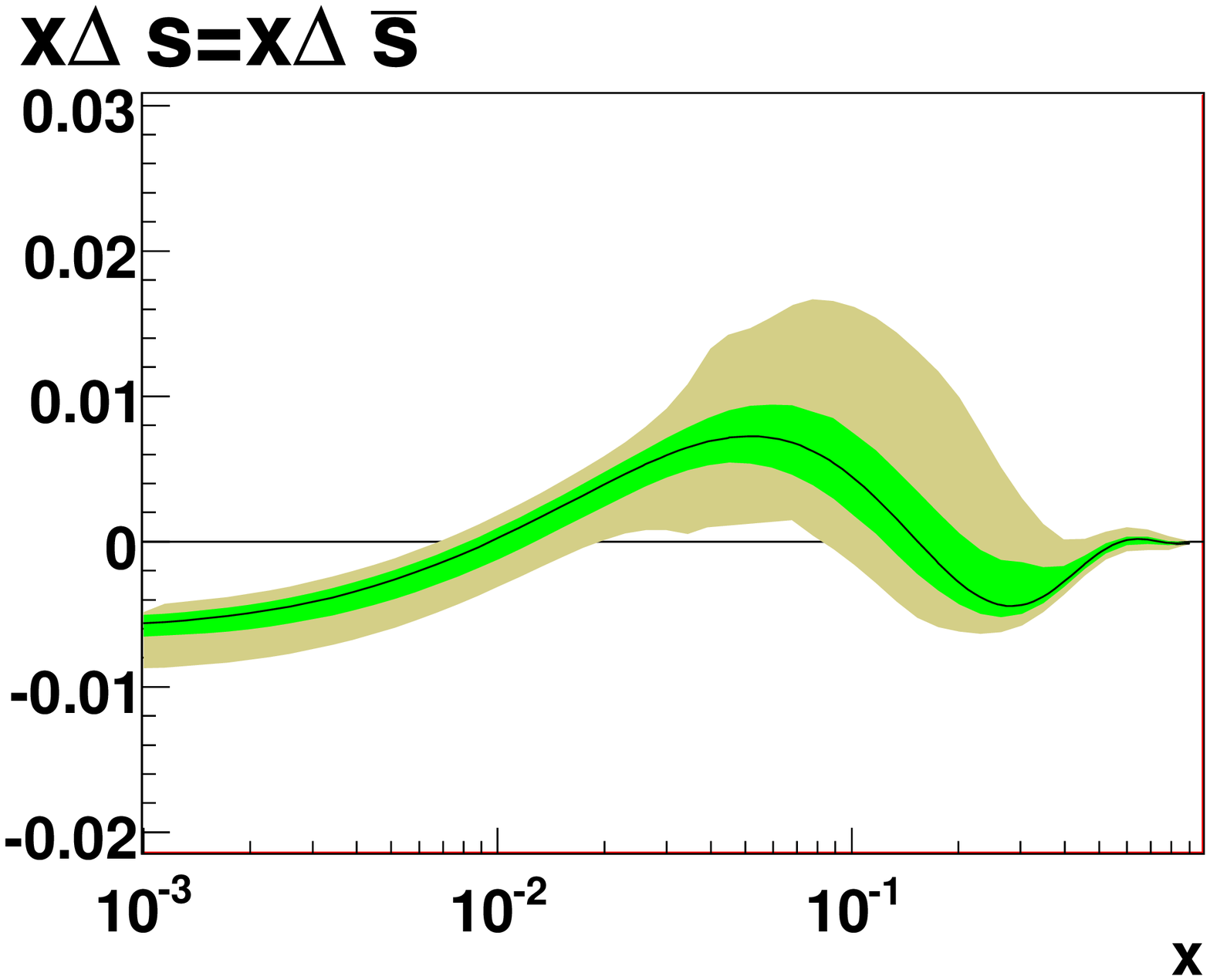}&
\includegraphics[height=5cm]{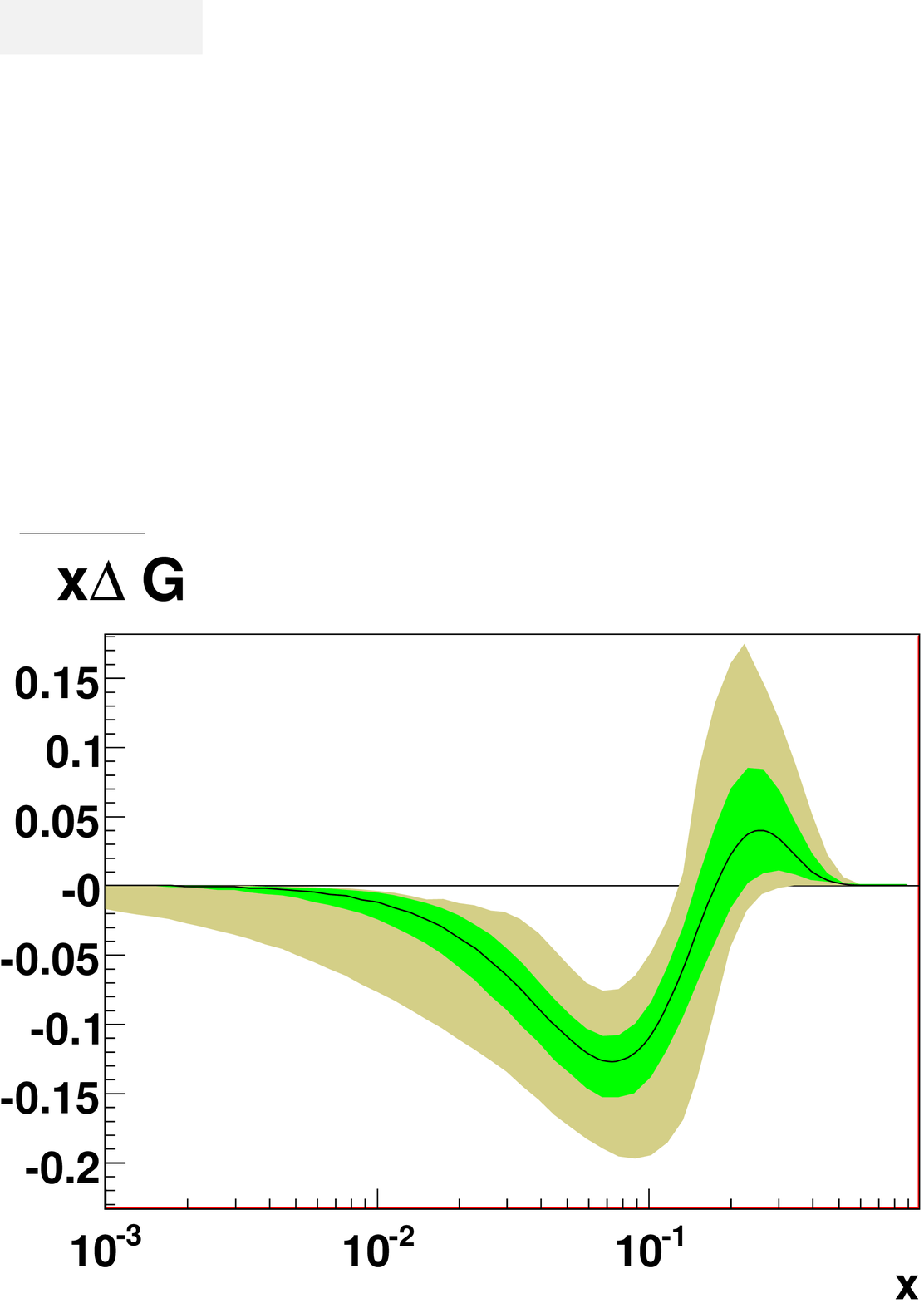}\\
\includegraphics[height=5cm]{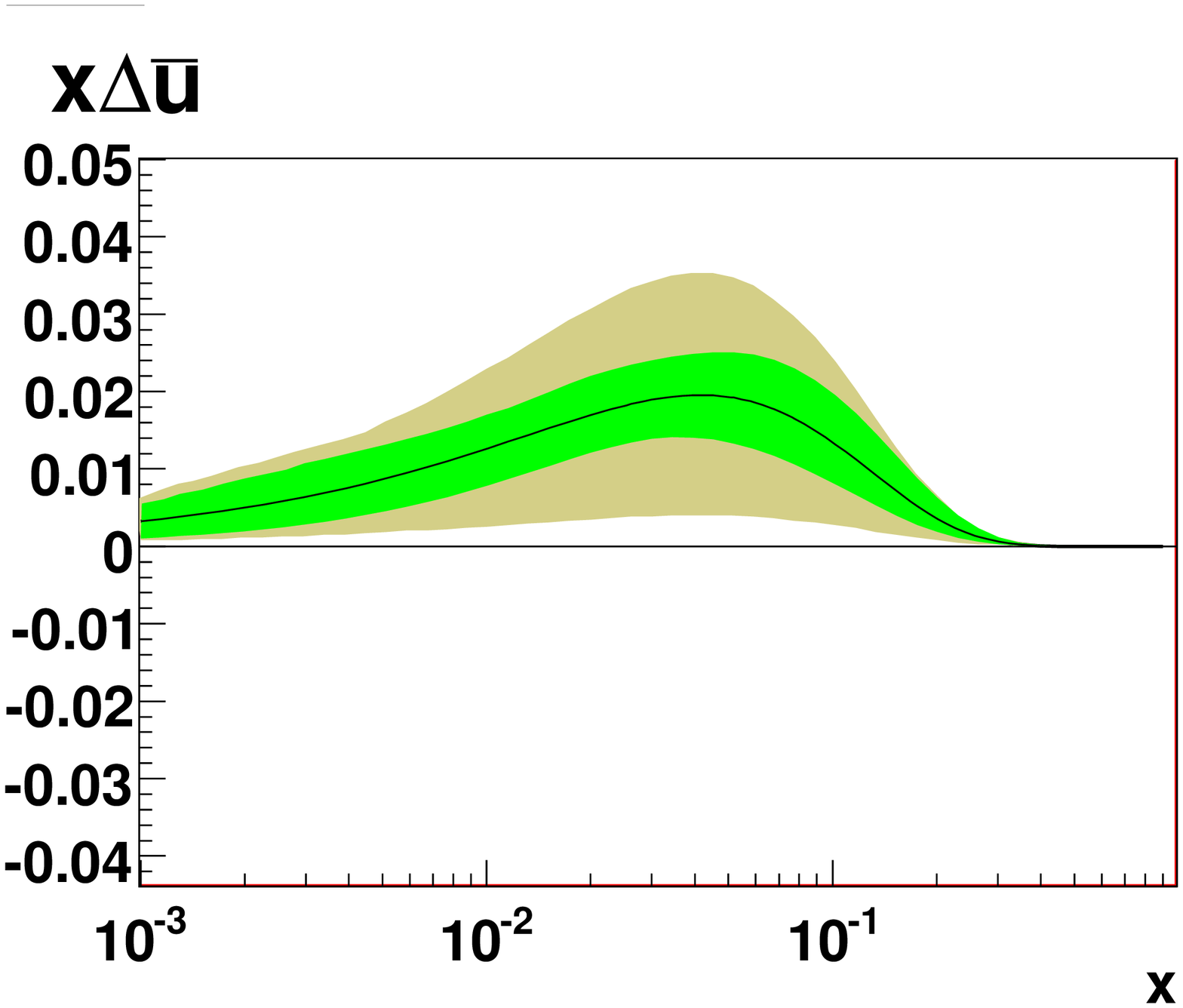}&
\includegraphics[height=5cm]{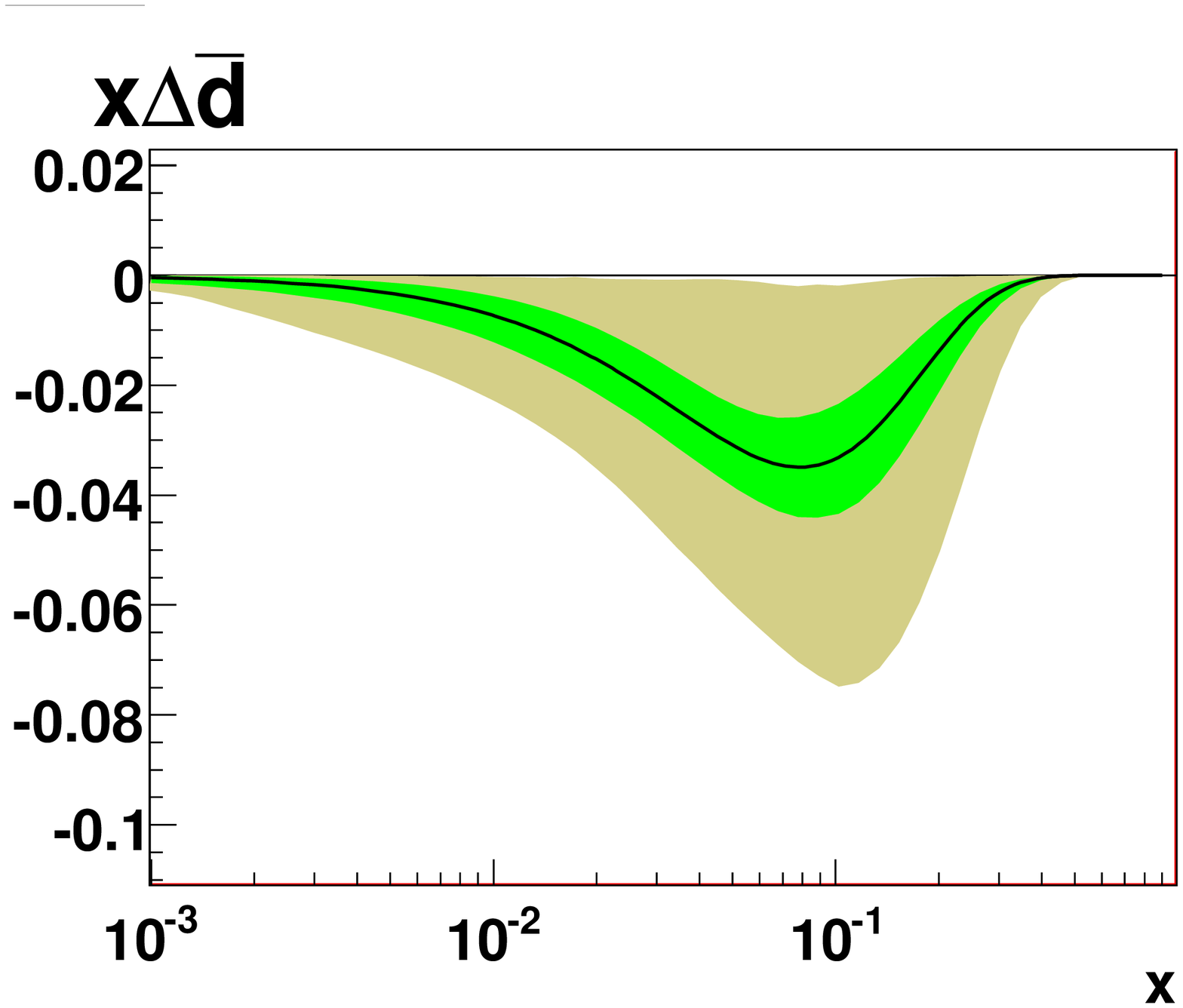}
\end{tabular}
\caption{Best fit PDFs (solid lines)
 together with their uncertainties for $\Delta\chi^2=1$ (inner bands) 
and $\Delta\chi^2=18.065$ (outer bands).}
\label{fig:local_dist}
\end{figure}

The local distributions together with their uncertainties are presented in Fig. \ref{fig:local_dist}. 
The  first moments of PDFs together with their uncertainties are presented in Table \ref{table:moments}.
As it was mentioned above, the results for both scenarios with fixed and 
varied  $\eta_{\Delta q_3}$ and $\eta_{\Delta q_8}$ are almost the same (the differences
in fitting parameters are less than 1\%). That is why we present all figures and Table \ref{table:moments}
 only for the option with the varied   $\eta_{\Delta q_3}$ and $\eta_{\Delta q_8}$.

Let us now discuss the obtained parametrization. First, one can see that the results on the first moments 
$\Delta_1\Sigma\equiv\eta_{\Delta\Sigma}$ and $\Delta_1 G\equiv\eta_{\Delta G}$ are very close to the respective results (scenario with
$\Delta G<0$) obtained in Ref. \cite{compass_dis} in the case of pure inclusive DIS. Indeed,
even with the more smaller errors, corresponding to the option $\Delta\chi^2=1$,
at $Q^2=3\,GeV^2$ we have $\Delta_1\Sigma=0.372^{+{0.008}}_{-0.014}$ 
and  $\Delta_1 G=-0.161^{+0.096}_{-0.146}$ in this paper and $\Delta_1\Sigma=0.329^{+0.004}_{-0.012}$
and  $\Delta_1 G=-0.181^{+0.042}_{-0.031}$ in Ref. \cite{compass_dis}, respectively. 
Notice that the obtained value $\Delta_1\Sigma=0.372^{+0.008}_{-0.014}$ is even more 
close to the respective NLO value $a_0(Q^2=3\,GeV^2)=0.35\pm 0.03$ obtained by COMPASS \cite{compass_dis} directly 
(without fitting procedure), from the first moment of the measured structure function $g_{1d}$.
From Table \ref{table:moments} one can see also that in the more pessimistic case $\Delta\chi^2=18.065$,
$\Delta_1 G$ value is just zero within the errors -- we will return to this question below. 
The impact of SIDIS data on the 
$\Delta G(x)$  shape will be also discussed some later.
What concerns nonsinglet combinations $\Delta q_3(x)$ and $\Delta q_8(x)$, we see (compare Table \ref{table:parameters}
with Table 3 from Ref. \cite{compass_dis}) that, even despite the small difference
in initial scales ($Q_0^2=1\,GeV^2$ here and $Q_0^2=3\,GeV^2$ in Ref. \cite{compass_dis}), the values
of $\Delta q_3(x)$ are very similar for both parametrizations. At the same time, the values of $\Delta q_8(x)$ 
differ much more essentially,
which occurs due to the impact (first of all on $\Delta s$ -- see below) of the additional SIDIS data.

\begin{table}
\caption{Estimations of the uncertainties on the first moments of polarized PDFs for
two options of $\Delta\chi^2$ choice.}
\begin{tabular}{|c|c|c|c|}
\hline
  &$\Delta\chi^2=1$&$\Delta\chi^2=18.065$\\
\hline
  $\Delta\Sigma$       &$ 0.3846^{+0.0050}_{-0.0122}$&$ 0.3846^{+0.0342}_{-0.0389}$\\
$\Delta u+\Delta\bar u$&$ 0.8640^{+0.0028}_{-0.0049}$&$ 0.8640^{+0.0114}_{-0.0084}$\\
$\Delta d+\Delta\bar d$&$-0.4020^{+0.0028}_{-0.0048}$&$-0.4020^{+0.0115}_{-0.0130}$\\
$\Delta s=\Delta\bar s$&$-0.0387^{+0.0014}_{-0.0024}$&$-0.038738^{+0.0061}_{-0.0065}$\\
$\Delta G$             &$-0.1828^{+0.0720 }_{-0.1090}$&$-0.1828^{+0.1693}_{-0.2831}$\\
$\Delta\bar u$         &$ 0.0672^{+0.0263}_{-0.0270}$&$ 0.0672^{+0.06483}_{-0.0737}$\\
$\Delta\bar d$         &$-0.0792^{+0.0191}_{-0.0238}$&$-0.0792^{+0.0795}_{-0.0830}$\\
\hline
\end{tabular}
\label{table:moments}
\end{table}

Let us now compare our results on $\Delta G$ with the respective results 
from Refs. \cite{dssv08} and \cite{aac08}.  The such comparison seems to be 
reasonable because among the number of all parametrizations applying the DIS data, the SIDIS data
is included only in the sequel of papers Ref. \cite{florian00}, \cite{florian05}, \cite{dssv08}. At the same time, though 
SIDIS data is not included in the analysis of Ref. \cite{aac08}, the RHIC $\pi^0$
production data is added\footnote{In this paper we are mainly interested in
the study of the SIDIS data impact on the polarized PDFs 
(especially on the still poorly known sea quark PDFs). 
The influence of RHIC $\pi^0$ production data on our parametrization will be considered
some later.
} there (just as in Ref. \cite{dssv08}), which should provide significant impact on $\Delta G$ 
in the RHIC $x$ region $[0.05,0.2]$. 
\begin{figure}[h!]
\includegraphics[height=5cm]{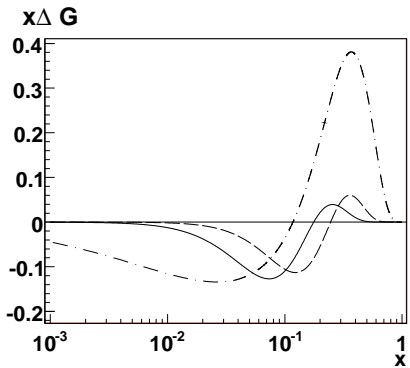}
\caption{Obtained parametrization on $\Delta G$
(solid line) in comparison with the respective parametrizations from Ref. \cite{dssv08} (dashed line)
and Ref. \cite{aac08} (dot-dashed line) at $Q_0^2=1\,GeV^2$.}
\label{fig:sigma_our_dssv}
\end{figure}
The results of comparison are presented in Fig. \ref{fig:sigma_our_dssv}.
From this figure one can see that the best fit results on $\Delta G$ presented in this paper
and in Ref. \cite{dssv08} are very close to each other while they are significantly differ
(in both $\Delta G<0$ and $\Delta G>0$ regions) from the respective result of 
Ref. \cite{aac08}, even for the same sign-changing scenario
for $\Delta G$. On the other hand, comparing the results
on the first moment $\Delta_1 G\equiv \eta_G$,
one can see that the central values of this quantity
are almost the same for all three parametrizations: these are $-0.183$, $-0.118$ and $-0.120$ in this paper,
Refs. \cite{dssv08} and \cite{aac08}, respectively. Moreover, including the uncertainties in comparison
(see Table \ref{table:moments} here, Table III in Ref. \cite{dssv08} and Table IV in Ref. \cite{aac08}),
we see  that the first moment $\Delta_1 G$ is just zero within the errors 
for all three parametrizations. Thus, we can conclude that with the present quality of the data
it is hardly possible to realize is the first moment $\Delta_1 G$ zero or not. To answer this question
and also to distinguish between the different shapes of $\Delta G(x)$, still more\footnote{Notice
that within this paper we apply the sign-changing scenario for
$\Delta G(x)$ and do not consider scenario $\Delta G(x)>0$ which was considered
in Refs. \cite{compass_dis} and \cite{aac08}. For this scenario the first moment
of $\Delta G$ is of opposite sign and larger in absolute value, about $+0.3\div+0.4$
instead of $-0.1\div-0.2$. Nevertheless,
the such scenario also produce quite acceptable $\chi_0^2$ value. The such arbitrariness even in the sign of
$\Delta_1 G$ once again tell us that we still need more data to properly fix $\Delta G$ and 
that even inclusion of $\pi^0$ production data (Refs. \cite{dssv08}, \cite{aac08}) still poorly 
enables us to solve the problem.} precise data is necessary.

\begin{figure}
\begin{tabular}{cc}
\includegraphics[height=5cm]{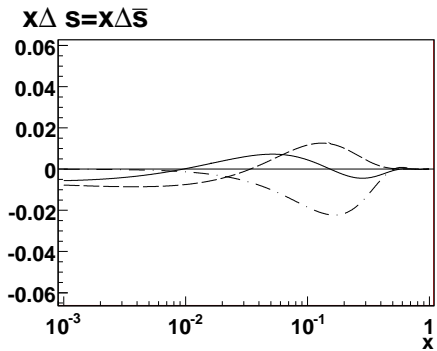}
\end{tabular}
\caption{Obtained parametrization on $\Delta s$ (solid line)
in comparison with the respective parametrizations from Ref. \cite{dssv08} (dashed line)
and \cite{compass_dis} (dot-dashed line), $Q^2=1\,GeV^2$.}
\label{ds_my_dssv_1}
\end{figure}

\begin{figure}
\begin{tabular}{cc}
\includegraphics[height=5cm]{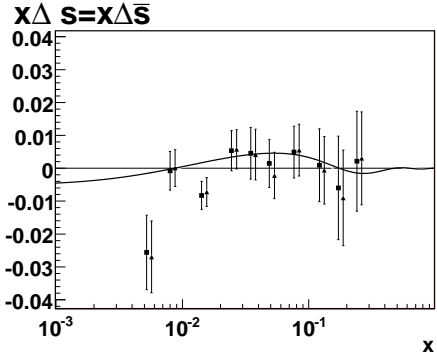}
\end{tabular}
\caption{Obtained NLO parametrization on $\Delta s $ (solid line)
in comparison with the COMPASS results \cite{compass09}  obtained in LO QCD (points
with error bars), $Q^2=3\,GeV^2$.}
\label{ds_my_compass}
\end{figure}

Let us now discuss the impact of SIDIS data on the polarized strangeness in nucleon.
It is of importance because, after the appearance of the first results on $\Delta s$
extraction from SIDIS data performed by HERMES \cite{hermes04},
we met the puzzle with the positive $\Delta s$ in the middle $x$ HERMES region $0.023<x<0.6$,
while the total moment $\Delta_1 s$ definitely should be negative \cite{leader-stamenov} in accordance with the 
sum rule (\ref{eq8}). Thus, to satisfy this requirement, $\Delta s(x)$ should possess the 
compensating negative behavior in the unaccessible for HERMES low $x$ region $0<x<0.023$,
i.e. the sign-changing scenario for $\Delta s$ should be realized.
Looking at Fig. \ref{fig:local_dist} we see that this is indeed the case and we produce the best
fit namely for the sign-changing $\Delta s$ scenario, as well as in Ref. \cite{dssv08}.
However, we see also the distinction in the $\Delta s$ shape, in comparison with Ref. \cite{dssv08} -- 
see Fig. \ref{ds_my_dssv_1}. 
Namely, while within parametrization \cite{dssv08} $\Delta s$ changes 
the sign one time, within our parametrization $\Delta s$ changes the sign twice. It seems 
that this distinction occurs due to the inclusion of the latest COMPASS semi-inclusive data \cite{compass09}, 
which allow to better fix\footnote{
Notice that after inclusion of the latest COMPASS data to our fit $\chi^2/D.O.F.$ value becomes small only
if we allow $\Delta s$ to change the sign twice (due to the additional parameter $\delta_{\Delta q_{8}}\sqrt{x}$ ).} 
$\Delta s$ shape.
%
This is illustrative to compare the NLO results on $\Delta s$ obtained here with the 
respective results of direct $\Delta s$ extraction in LO by COMPASS -- see Fig. \ref{ds_my_compass}.
We see very similar $\Delta s$ behavior in both cases. Certainly, one should be careful comparing
LO and NLO results. However, the LO and NLO results do not differ too drastically, so that
the such comparison is very useful and allow us to make
at least qualitative conclusion about the PDFs behavior (shape of distributions).
The new COMPASS data on the kaon asymmetries in the wide Bjorken $x$ region $[0.003,0.7]$ 
will be available in the nearest future (the paper in preparation). Thus, one can hope to eventually
fix the strangeness in nucleon.

\begin{figure}[h!]
\includegraphics[height=5cm]{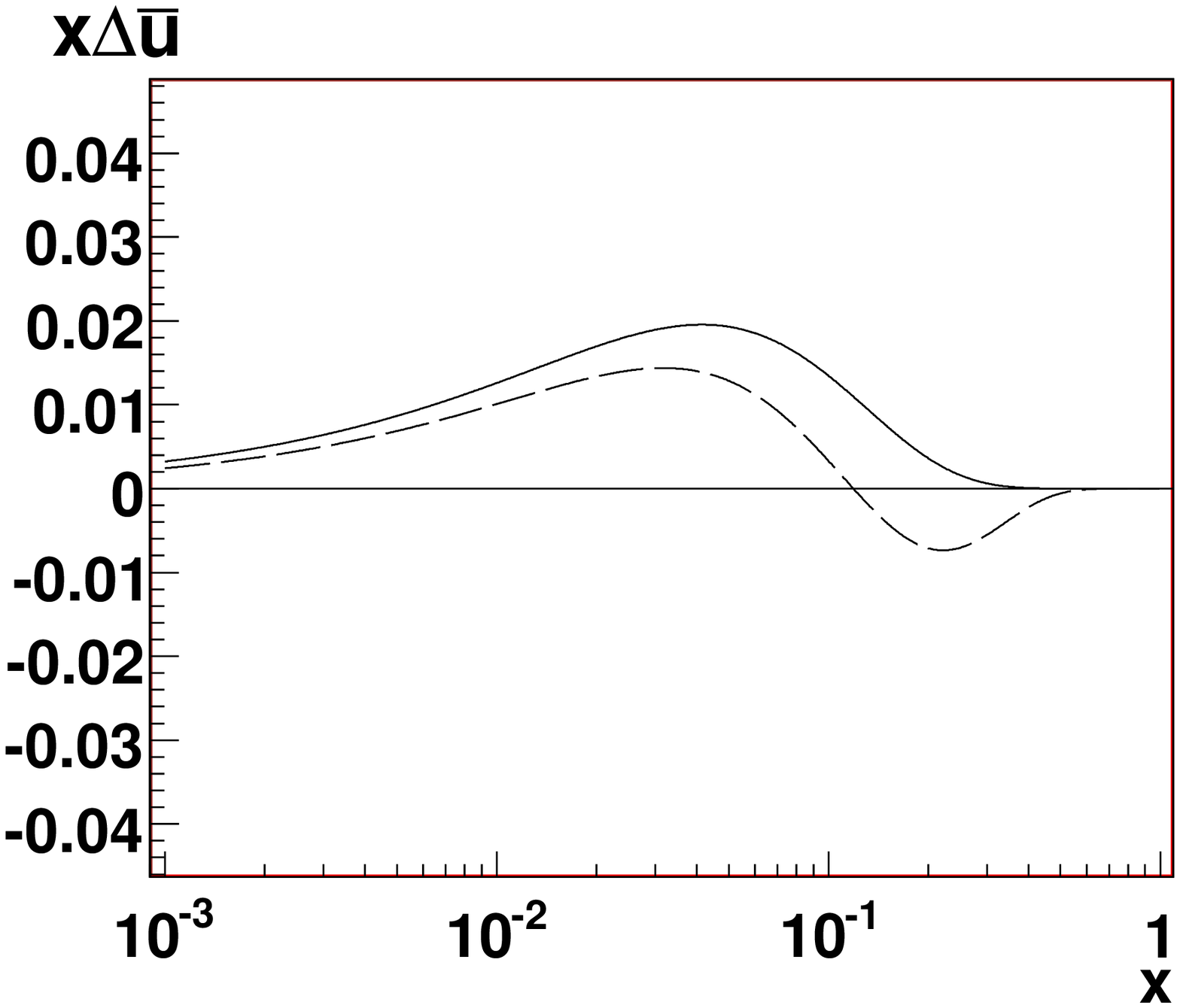}
\includegraphics[height=5cm]{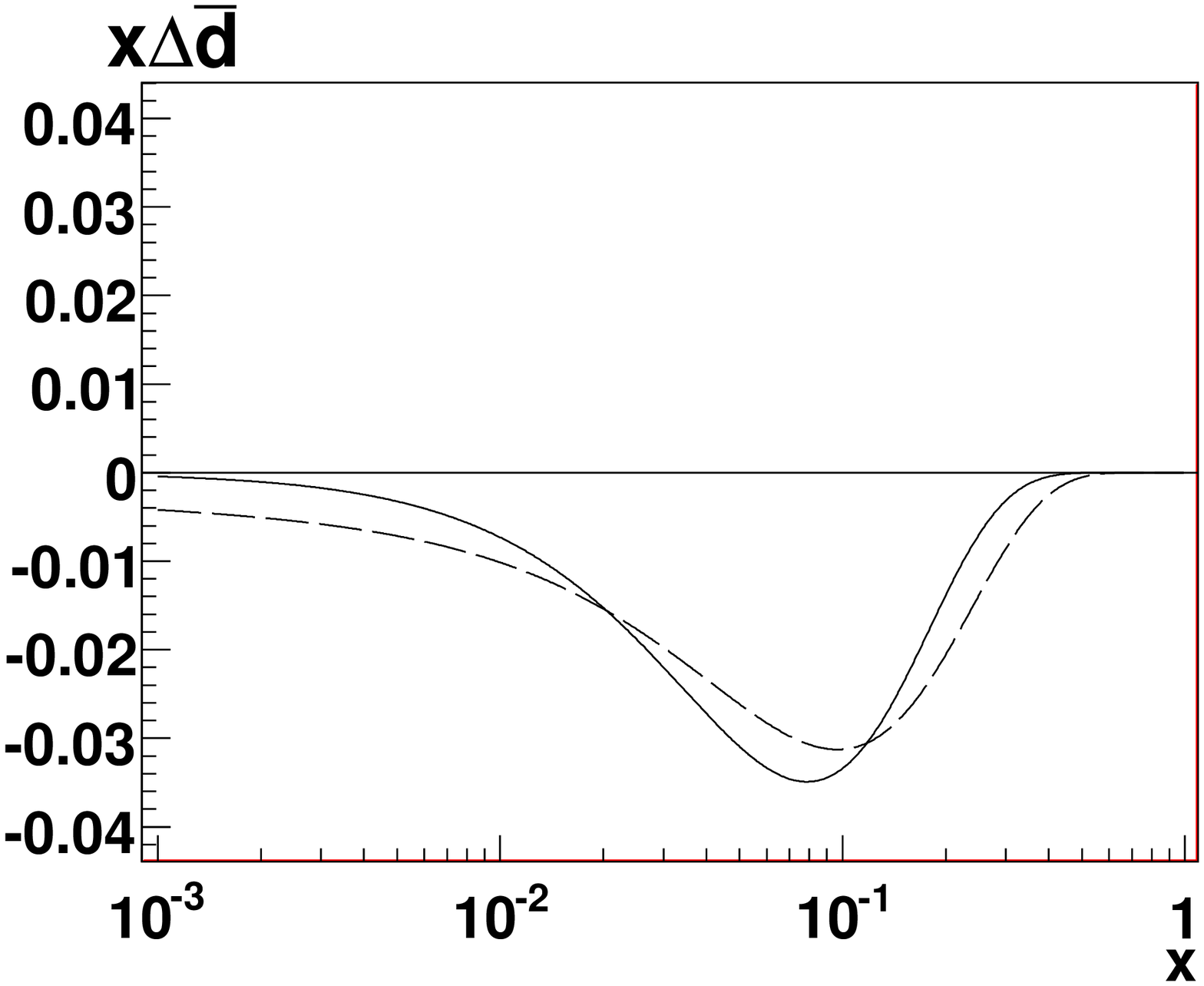}
\caption{Obtained parametrization on $\Delta\bar u$ and $\Delta\bar d$ (solid lines)
in comparison with the respective parametrization of Ref. \cite{dssv08} (dashed lines), $Q^2=1\,GeV^2$.}
\label{ubar_dbar_our_dssv}
\end{figure}

\begin{figure}
\includegraphics[height=5cm]{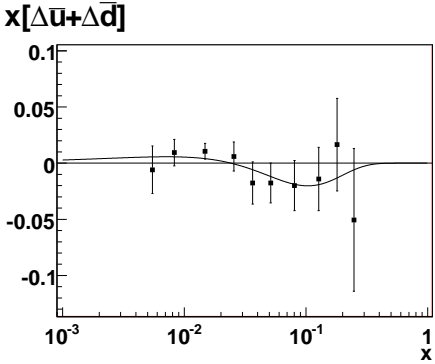}
\caption{Obtained NLO parametrization on $\Delta\bar u+\Delta\bar d$  (solid line)
in comparison with the COMPASS results \cite{compass09}  obtained in LO QCD (points
with error bars), $Q^2=3\,GeV^2$}
\label{fig:ubdb}
\end{figure}

As it was discussed above, the exclusive task for SIDIS is the finding of sea $\Delta\bar q$ (and,
consequently, valence $\Delta q_V=\Delta q-\Delta\bar q$) PDFs in nucleon.
We again compare our results on $\Delta\bar u$ and $\Delta\bar d$ with the respective results from
Ref. \cite{dssv08}. Comparing the central values (best fit values) of $\Delta\bar u$ and $\Delta\bar d$
one can see that they are quite similar -- see Fig. \ref{ubar_dbar_our_dssv}. At the same time, there 
are also some distinctions, and main of them are connected with the sum $\Delta\bar u(x)+\Delta\bar d(x)$ and its first
moment $\Delta_1\bar u+\Delta_1\bar d$. 
The point is that recently, analyzing the SIDIS data
on $h^\pm$ production COMPASS obtained rather surprising result \cite{compass_hpm} that
the sum  $[\Delta_1\bar u+\Delta_1\bar d](Q^2=10\,GeV^2)$ is just zero within the errors (see Table 2 in Ref. \cite{compass_hpm})
\be
\label{eq_comp_ubdb}
[\Delta_1\bar u+\Delta_1\bar d]_{COMPASS}=0.0\pm0.04\pm0.03.
\ee
This result  was confirmed in the subsequent COMPASS paper \cite{compass09}, where 
sum $\Delta\bar u(x,Q^2=3\,GeV^2)+\Delta\bar d(x,Q^2=3\,GeV^2)$ of the local PDFs was extracted from the measured asymmetries
$A_{1d},A_{1d}^{\pi^\pm}, A_{1d}^{K^\pm}$ in the region $0.004<x<0.3$ 
(see Fig. 4 in Ref. \cite{compass09}) and occurs to be about zero in the whole this region (the central 
values occur in both positive and negative vicinities of zero).
 Thus, at least in the leading order (COMPASS analysis) the sum $\Delta\bar u(x)+\Delta\bar d(x)$
is about zero in the region $3\,GeV^2<Q^2<10\,GeV^2$, which sheds new light on our understanding of polarized light
quark sea. 
Namely, the sea is extremely asymmetric ($\Delta\bar u\simeq-\Delta\bar d$), on the contrary to the assumption of
symmetric sea scenario 
$\Delta\bar u(x,Q_0^2)=\Delta\bar d(x,Q_0^2)$, applied in the practically all\footnote{\label{footnot} The only exception is asymmetric
GRSV2000 parametrization \cite{grsv2000}. However, though in this parametrization
$\Delta\bar u$ and $\Delta\bar d$ are also of opposite sign, they are strongly differ in absolute value.} existing parametrizations based on the pure
inclusive DIS data analysis.
Our analysis shows that the sum $\Delta\bar u+\Delta\bar d$ is very small quantity
in NLO QCD too. It is very illustrative for qualitative comparison to put our 
best NLO QCD fit  for $\Delta\bar u(x)+\Delta\bar d(x)$ (evolved to the COMPASS $Q^2=3\,GeV^2$) 
on the figure (Fig. 4 in Ref. \cite{compass09}) showing the
COMPASS LO results on this quantity -- see Fig. \ref{fig:ubdb}. From Fig. \ref{fig:ubdb} it is clearly seen that
 $\Delta\bar u(x)+\Delta\bar d(x)$ is very small quantity in both LO and NLO QCD orders. 
In turn, the first moment $\Delta_1\bar u+\Delta_1\bar d$ for the proposed parametrization
is also just zero within the errors
\be
\label{eq_our_ubdb}
[\Delta_1\bar u+\Delta_1\bar d](Q^2=1\,GeV^2)=-0.01^{+0.01}_{-0.02},
\ee
even in the case $\Delta\chi^2=1$. Notice that QCD evolution weakly influences this result. Even at 
extremely large $Q^2$ value 100 GeV$^2$ (for instance, for COMPASS upper bound on accessible $Q^2$
is about 60 GeV$^2$) the quantity $\Delta_1\bar u+\Delta_1\bar d$ is still very close to zero within the errors,
even in the case $\Delta\chi^2=1$:
\be
\label{eq_our_ubdb_100}
[\Delta_1\bar u+\Delta_1\bar d](Q^2=100\,GeV^2)=-0.02^{+0.01}_{-0.03}.
\ee

 Notice that looking on parametrization
of Ref. \cite{dssv08} (see Fig. 3 there) we see that the distributions $\Delta\bar u(x)$ and $\Delta\bar d(x)$
are also of opposite sign. However, they are significantly differ in their absolute values
(just as in the DIS parametrization GRSV2000 \cite{grsv2000} -- see footnote \ref{footnot}).
As a result, the central value of $[\Delta_1\bar u+\Delta_1\bar d](Q_0^2=1\,GeV^2)$  in Ref. \cite{dssv08} (see Table IV
there)  is $-0.08$ (i.e. almost the same as $\Delta_1\bar d $ 
central value $-0.11$) instead of $-0.01$ in Eq. (\ref{eq_our_ubdb}).

The important remark should be made here. At present, the SIDIS data is of such quality that all above conclusions 
about sea PDFs should be considered as preliminary. Indeed,  looking at Fig. \ref{fig:local_dist} here and Fig. 3 
in Ref. \cite{dssv08} we see that the uncertainties on $\Delta\bar u$ and $\Delta\bar d$ are rather large even for the 
option $\Delta\chi^2=1$, while in the case of $\Delta\chi^2$ determined by Eq. (\ref{chi2}) ($\Delta\chi^2=18.065$ here)
$\Delta\bar u$ and $\Delta\bar d$ are still the zeros within the errors, and even  just to see them 
within this option for $\Delta\chi^2$ we need more SIDIS data (first of all the expected COMPASS data).


{\it In conclusion}, the new combined analysis of polarized DIS and SIDIS data in NLO QCD is presented\footnote{The 
code is available from authors upon request.}. 
The impact of modern SIDIS data on polarized PDFs is studied, which is 
of especial importance for the light sea quark PDFs and strangeness in nucleon.
The obtained results are in agreement with the latest direct leading order
COMPASS analysis of SIDIS asymmetries \cite{compass09}
as well as with the recent global fit analysis in NLO QCD of  Ref. \cite{dssv08}, where
the SIDIS data were also applied. Nevertheless, there also some distinctions concerning, first of all, the polarized 
quark sea peculiarities. At the same time, the quality of SIDIS data is still not sufficient
to make the eventual conclusions about the quantities influenced mainly by SIDIS.
In this situation of especial importance becomes the direct $\Delta q$ extraction in NLO QCD,
where (just as in LO QCD) the central values of asymmetries and their uncertainties directly propagate
to the extracted $\Delta q$ values and their errors. The such NLO QCD method, free of any 
fitting procedure with a lot of parameters,  was elaborated in the sequel of papers \cite{method} 
(see review \cite{echaya} for details). 
At present the respective analysis of the whole existing polarized DIS and SIDIS data
is in preparation. In any case, the new COMPASS semi-inclusive data should be available in the nearest future.
In particular, we expect $\pi^\pm$, $K^\pm$ data on the proton target in the wide $x$ region (today the only available such
data is the HERMES data in the narrow $x$ region and only for $\pi^\pm$ production),
which should essentially increase the precision of $\Delta\bar u$, $\Delta\bar d$ and $\Delta s$
extraction.

The authors are grateful to Y. Bedfer, E. Kabuss, 
A. Kotzinian, A. Korzenev, F. Kunne, A. Nagaytsev, A. Magnon, M. Stratmann,
 G. Piragino, E. Rondio, I. Savin, H. Santos, A. Sidorov and 
 R. Windmolders for fruitful discussions.
The work of O. Shevchenko and O. Ivanov was supported
by the Russian Foundation for Basic Research (Project No. 07-02-01046).


\end{document}